\documentclass{article}
\usepackage[preprint]{neurips_2020}
\usepackage{geometry}
\usepackage{graphicx}
\usepackage{qtree}
\usepackage{tabularx}
\usepackage{booktabs}
\usepackage{siunitx}
\usepackage{ragged2e}
\usepackage{rotating}
\usepackage{adjustbox}
\usepackage{caption}
\usepackage[T1]{fontenc}
\usepackage{pdflscape}
\usepackage{nameref}
\usepackage{pdfpages}
\usepackage[multiple]{footmisc}
\usepackage{float}
\newcolumntype{R}[1]{>{\RaggedRight}p{#1}}
\usepackage{amssymb}
\usepackage{xcolor}
\usepackage{verbatim}
\usepackage{subcaption}
\usepackage{array}
\usepackage{natbib}
\usepackage{hyperref}
\usepackage{mathtools}
\usepackage{xspace}
\usepackage{cleveref}
\usepackage{svg}
\usepackage{transparent}
\usepackage[toc,title,page]{appendix}
\definecolor{orangefull}{RGB}{230, 159, 0}
\newcommand*{\eg}{e.g.\@\xspace}
\newcommand*{\ie}{i.e.\@\xspace}

\usepackage{siunitx}
\usepackage[pagewise]{lineno}  
\colorlet{orange}{orangefull!20!white}
\definecolor{bluefull}{RGB}{86, 180, 233}
\colorlet{blue}{bluefull!20!white}
\definecolor{green}{RGB}{0, 158, 115}
\colorlet{green}{green!20!white}

\title{Bridging CORDEX and CMIP6: Machine Learning Downscaling for Wind and Solar Energy Droughts in Central Europe}

\author{%
  Nina Effenberger \\
  Institute for Atmospheric and Climate Science\\
  ETH Zurich\\
  \texttt{nina.effenberger@env.ethz.ch} \\
  \And
  Maxim Samarin \\
  Swiss Data Science Center\\
  EPFL and ETH Zurich\\
  \And
  Maybritt Schillinger \\
  Seminar for Statistics\\
  ETH Zurich\\
  \And
  Reto Knutti \\
  Institute for Atmospheric and Climate Science\\
  ETH Zurich
}
  
\begin{document}

\maketitle

\begin{abstract}
Reliable regional climate information is essential for assessing the impacts of climate change and for planning in sectors such as renewable energy; yet, producing high-resolution projections through coordinated initiatives like CORDEX that run multiple physical regional climate models is both computationally demanding and difficult to organize. Machine learning emulators that learn the mapping between global and regional climate fields offer a promising way to address these limitations. Here we introduce the application of such an emulator: trained on CMIP5 and CORDEX simulations, it reproduces regional climate model data with sufficient accuracy. When applied to CMIP6 simulations not seen during training, it also produces realistic results, indicating stable performance. Using CORDEX data, CMIP5 and CMIP6 simulations, as well as regional data generated by two machine learning models, we analyze the co-occurrence of low wind speed and low solar radiation and find indications that the number of such energy drought days is likely to decrease in the future. Our results highlight that downscaling with machine learning emulators provides an efficient complement to efforts such as CORDEX, supplying the higher-resolution information required for impact assessments.
\end{abstract}

\section{Introduction}
Global or regional climate models provide the foundation for assessing future changes in weather and climate. The latest generation of global projections, CMIP6 \citep{eyring2016overview}, improves upon CMIP5 in several aspects, including the representation of wind \citep{carvalho2021wind} and solar radiation fields \citep{ha2023comparative}. However, because high-resolution regional CMIP6-based data are not yet available, impact studies often face a trade-off between using the most up-to-date coarse-resolution CMIP6 projections and higher-resolution regional CORDEX data \citep{giorgi2015regional} from CMIP6's predecessor, CMIP5 \citep{rummukainen2016added, gibba2019state}. Therefore, there is a need for approaches that can harness the improved global CMIP6 information while retaining regional detail, especially for impact studies such as those related to renewable energy.
The transition to renewable energy sources is widely recognized as a central element of future energy systems; yet, the variable nature of wind and solar power poses a fundamental challenge to energy security \citep[\eg][]{heylen2018review}. Particularly, periods of low renewable generation due to low wind speeds and/or low solar radiation, often referred to as \textit{energy droughts} or \textit{Dunkelflaute} days \citep{kittel2024measuring}, can place severe stress on electricity systems that rely on variable renewables \citep[\eg][]{kapica2024potential}. Such compound events are more critical than single-resource deficits, as they restrict the balancing potential between wind and solar power \citep{kittel2024coping, li2024renewable, bracken2024standardized}. Analyzing how such events may evolve under future climate conditions is, therefore, essential for resilient energy system planning \citep[\eg][]{xie2024role}.

Machine learning (ML) emulators offer a potential solution by learning the relationships between coarse-resolution global climate models (GCMs) and high-resolution regional climate models (RCMs), approximating RCM behavior at a fraction of the computational cost \citep{rampal2024enhancing}. This process is usually referred to as ML downscaling and derives fine-scale local climate information from coarse GCM outputs. Traditional downscaling approaches include dynamical downscaling, which uses high-resolution RCMs nested within GCMs, and statistical downscaling, which links large-scale model outputs to local-scale variables \citep{ekstrom2015appraisal}. ML emulators can be trained directly on RCM outputs, thereby linking physical modeling and statistical approaches \citep{van2023deep}. However, the development of such emulators still requires high-resolution RCM simulations as training data. At the same time, their integration into applied impact studies remains limited, with practical uptake lagging behind methodological progress \citep{fowler2025downscaling}. A major challenge lies in establishing whether emulators can provide reliable results beyond their training conditions \citep{hernanz2024limitations} and in demonstrating their usefulness for impact-relevant tasks \citep{fowler2025downscaling, fowler2007linking}. However, there is work bridging this gap: \citet{strnad2025assessingriskfuturedunkelflaute} recently applied a generative deep learning framework to downscale CMIP6 climate simulations, assessing future Dunkelflaute events in Germany under various emission scenarios, and \citet{lewis2025generative} investigating changes in future drought events.

In this study, we train a machine learning emulator on coupled CMIP5-CORDEX simulations, validate its performance on an impact-focused task, and then apply it to CMIP6 data. Focusing on energy droughts, we analyze the occurrence of days with simultaneously low wind and solar radiation in multi-model ensembles \citep{jung2022review, kapica2024potential}. The probabilistic nature of the used ML emulator enables efficient exploration of inter-model spread, providing a way to assess uncertainty and extremes across multiple climate realizations. To our knowledge, this is one of the first applications of an emulator trained on CMIP5 and CORDEX, applied to CMIP6 projections, and the first to specifically extend its use to energy droughts. Importantly, we present both validation and application, providing new insights into future risks for renewable energy systems. Further, we demonstrate that the gap between methodological development and applied climate impact research can be bridged. 

\section{Methods}
Our workflow involves training a machine learning emulator on coupled CMIP5-CORDEX data, evaluating its performance on energy droughts, and subsequently applying it to CMIP6 projections. In this section, we describe the datasets and preprocessing steps, the design and training of the machine learning model, our definition of energy droughts, and the analyses performed.

\label{methods}
\subsection{Data}
For our analyses, we mainly use data from CMIP5 \citep{taylor2012overview}, CMIP6 \citep{eyring2016overview}, and CORDEX \citep{jacob2014euro}. 
\Cref{tab:models} summarizes the CMIP5 and CORDEX global and regional climate models used in this study.  The variables selected are daily averages of surface wind speed (\textit{sfcWind}) and solar radiation (\textit{rsds}).
We focus on Central Europe, selecting the region bounded by longitudes $[-4.90, 19.18]$ and latitudes $[40.44, 55.97]$ from both RCM and emulated RCM data, as illustrated in \Cref{fig:location}. In our study, the term \textit{emulated data} refers to data generated by a machine learning model trained to reproduce RCM output based on coarser GCM input. Both RCM and emulated RCM data are provided on the same rotated grid (EURO-CORDEX EUR-11 grid with an approximate $0.11^\circ$ degree resolution). GCM data are on a regular latitude–longitude grid ($2.5^\circ$ resolution), which is consistent across all GCMs. Thus, the datasets differ not only in spatial resolution but also in grid orientation. We analyze two time periods, 2030–2039 and 2090–2099, under the RCP8.5 scenario for CMIP5 and its CMIP6 equivalent, SSP5-8.5. For CMIP6, the trained emulator is applied to the models CNRM-CM6-1 \citep{seferian2019evaluation}, MPI-ESM1-2-LR \citep{MPI-LR}, and MIROC6 \citep{tatebe2019description}. Since daily averaged surface wind data are not provided for CNRM-CM6-1, we compute them by aggregating the available 6-hourly averages. We also compare the results to ERA5-Land reanalysis data \citep{munoz2021era5}, using the period 2014–2023. In the ERA5-Land dataset, we use the two 10 m wind components on an hourly scale, \textit{u10} and \textit{v10}, to compute the wind speed as
\[
\mathit{sfcWind} = \sqrt{\text{u10}^2 + \text{v10}^2},
\]
and we use the surface solar radiation downward (\textit{ssrd}) variable. ERA5-Land provides hourly data, which we aggregate to daily averages for easier comparison. Although the temporal period and temporal resolution do not align with the future time horizons analyzed, ERA5-land serves as a reference point to contextualize the magnitude of observed drought events.

\begin{table}[h!]
\centering
\small
\caption{RCM–GCM combinations in the CMIP5 dataset. A check mark indicates that the combination exists and was used for training the machine learning model and in the data analysis presented here.}
\label{tab:models}
\begin{tabular}{lccc}
\hline
RCM / GCM & CNRM-CM5 & MIROC5 & MPI-ESM1-2-LR \\
\hline
ALADIN63    & \checkmark &          & \checkmark    \\
CCLM4-8-17  & \checkmark & \checkmark & \checkmark   \\
RegCM4-6    & \checkmark &          & \checkmark   \\
REMO2015    &            & \checkmark &               \\
\hline
\end{tabular}
\end{table}

\subsection{Downscaling emulator}
\paragraph{Data for training} Training the machine learning model is based on three CMIP5 global models and four EURO-CORDEX regional models (see \Cref{tab:models}) and follows the setup in \cite{schillinger2025enscaletemporallyconsistentmultivariategenerative}.
We consider daily data for the years 1971-2099, from both the historical and the RCP8.5 experiment. 1971-2029 and 2040-2089 serve as the training periods. 2030-2039 and 2090-2099 are used for testing purposes in \citet{schillinger2025enscaletemporallyconsistentmultivariategenerative} and are analyzed here.
During training, we consider daily 2 m temperature (\textit{tas}), total precipitation (\textit{pr}), surface wind (\textit{sfcWind}), and surface downwelling shortwave (solar) radiation (\textit{rsds}) from both, the driving GCMs and the corresponding downscaled RCMs. In addition, we also add sea level pressure (\textit{psl}) from the GCMs as a predictor.

\paragraph{ML model} 
\label{paragraph:ml_model}
As the machine learning model we employ \textit{CorrDiff} which was previously shown to successfully perform multivariate downscaling \citep{mardani2025residual, schillinger2025enscaletemporallyconsistentmultivariategenerative}. The full model description is provided in \citet{mardani2025residual}. 
Different from the original setting, we use a smaller U-Net in both the regression and diffusion parts (see \Cref{sec:additional_plots_spatial_avg}). In particular, we use a base embedding size of 64, multiply channels by the factors [1, 2, 2], and use no attention layers in the bottleneck representation. We found that for the energy drought application, this reduced setting provided more consistent results than the full setting reported in \citet{mardani2025residual}, see \Cref{fig:attention-emulator-cmip5}. We use a smaller batch size of 64, and train until convergence: first, the U-Net for 4 million training samples and, afterwards, the diffusion model for 24 million training samples. During inference, we generate 10 samples per model unless stated otherwise. For one RCM-GCM pair during our 10-year test periods, generating these samples takes an average of 4 hours on a NVIDIA A100 GPU.

\paragraph{Robustness of our results}
In addition to \textit{CorrDiff}, we test the ML model \textit{EnScale} described in \citet{schillinger2025enscaletemporallyconsistentmultivariategenerative} on the same CMIP5–CORDEX training dataset. This alternative model produced qualitatively similar results; therefore, we focus the main analyses on \textit{CorrDiff} and show the results of \textit{EnScale} in \Cref{sec:enscale}.

\subsection{Energy droughts}
Our definition of energy drought events follows \citet{kapica2024potential}. As in their study, we adopt a seasonal approach, dividing the year into four seasons: December, January, and February (winter in our study region); March, April, and May (spring); June, July, and August (summer); and September, October, and November (autumn). To identify critical low-renewable days, we first compute the daily spatial averages of wind speed \textit{sfcWind} and solar radiation \textit{rsds} across the region and then flag days as drought days when both averages simultaneously fall below a pre-defined threshold $p$. Following \citet{kapica2024potential}, we set $p$ at the 20th percentiles of wind speed and solar radiation, respectively. 
The drought index $d$ per day is then computed as
\begin{align}
d &= 
\begin{cases} 
1, & \text{if } sfcWind \leq p
\text{ and } rsds \leq p \\[2mm]
0, & \text{otherwise} 
\end{cases}
\label{eq:threshold}
\end{align}
where $p$ is computed separately for each dataset (GCM, RCM, or emulated RCM) and for each time period. Our drought-day definition thus acts as a compound metric capturing the co-occurrence of low wind and low solar radiation events. The total number of drought days per season is defined as the count of days in a given season when wind speed and solar radiation simultaneously fall within the lowest 20 \% of their respective distributions. This thresholding implicitly performs a quantile-mapping–like standardization, removing the need for an additional bias-correction step. We also investigate the spatial patterns of drought events, where we compute $p$ and the relative amount of drought days at each grid point. 

We define the threshold for energy droughts over the entire year rather than by season, as done in \citet{kapica2024potential}. This choice is motivated by research on Dunkelflaute events, which predominantly occur in autumn and winter \citep{mockert2022meteorological}, and ensures that periods potentially critical to the energy system are properly captured during the correct season instead of being equally distributed across the seasons. Furthermore, we use the raw wind speed and solar radiation data rather than capacity factors, which reduces the number of assumptions required. We discuss the advantages and limitations of this choice in \Cref{sec:discussion}. 

\paragraph{Change in drought days}
To assess future changes in energy droughts, we compare the number of drought days between the periods 2030–2039 and 2090–2099 for each season across the entire region. We then compute the mean and standard deviation of the drought days and changes thereof across the model ensembles, namely across all GCMs, all RCMs and all emulated RCMs, respectively. Note that we treat all three groups the same, with GCMs and RCM consisting of eight single members and the emulated RCM group always having 10 members per emulated RCM.  In addition, we quantify the agreement on the sign of change, defined as the fraction of ensemble members that consistently indicate an increase or decrease in drought days. This approach has been used in IPCC assessments to summarize the change in multi-model projections \citep{lee2021future}. 

\paragraph{Sampling extreme events}
Additionally, we assess whether the emulator can generate extremes beyond those present in the CORDEX training data and examine how this ability scales with the number of samples. To do so, we draw 50 samples per RCM using CMIP5 boundary conditions and compare these results with both the original CORDEX RCM data and the 10 emulator samples used for the other analyses.

\section{Results}
Our results include several analyses. First, we validate the trained emulator against CORDEX data to assess its ability to reproduce drought days in our study region. We also analyze the original CORDEX data to provide a baseline understanding of the evolution of drought days. To quantify differences in the spatial patterns of GCM and RCM data, we further examine the complete spatial fields. Second, we apply the emulator to CMIP6 data, allowing us to examine projected changes in drought characteristics at RCM-resolution. Finally, we assess the capability of the emulator to generate extremes beyond its training distribution. Together, these analyses provide a comprehensive assessment of both the performance of the emulator and the potential impacts of future climate change on energy-relevant drought events.

\subsection{Emulator validation and analysis of CORDEX data}
To quantify changes in co-occurring low wind and solar days, we compute the number of drought days per season over 10-year periods, as described in \Cref{methods}. We analyze these events using the original RCM simulations, the emulated RCM data, and the GCM data from two future periods, 2030–2039 and 2090–2099. For each dataset, we calculate the seasonal mean and standard deviation of drought days across ensemble members, allowing for a comparison of both the magnitude and consistency of the projected changes. 

\begin{figure}[H]
    \centering
    \includegraphics[width=\linewidth]{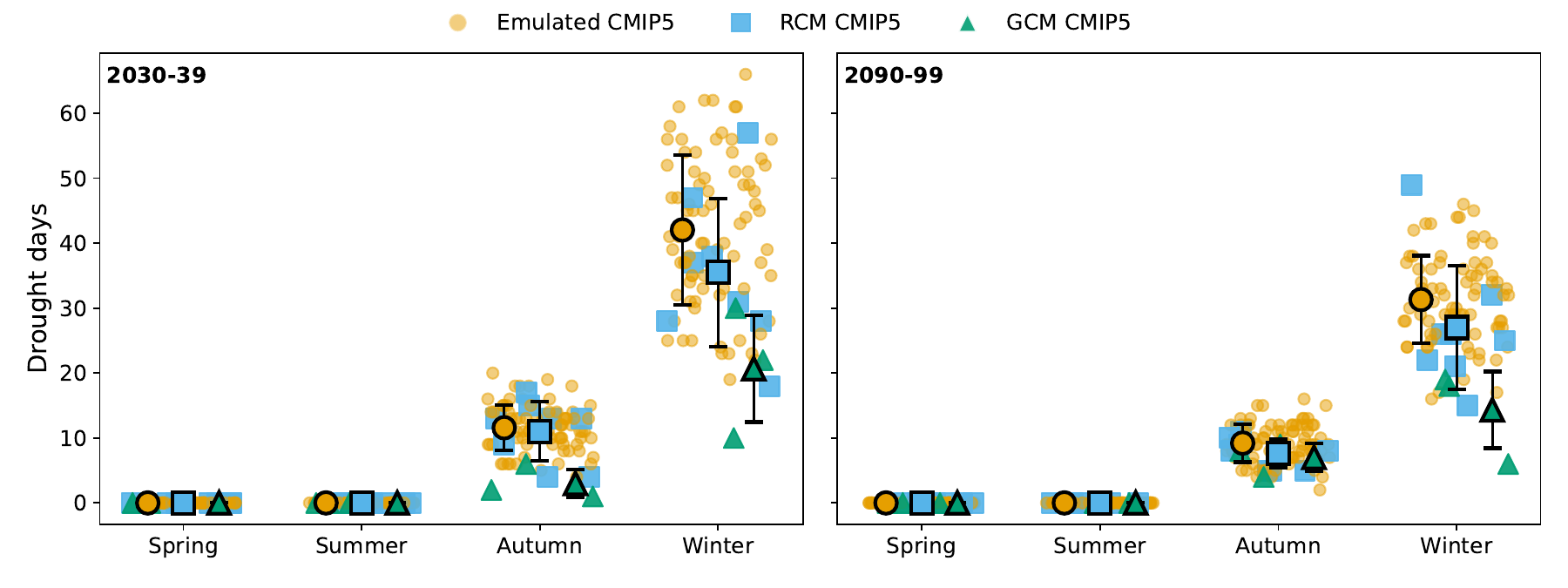}
    \caption{Number of co-occurring low wind and solar radiation days in the original RCMs from CORDEX, emulated RCMs, and GCMs from CMIP5 for the periods 2030–2039 (left) and 2090–2099 (right). No drought days are observed in spring and summer. In autumn and winter, both the emulated and original RCMs indicate a decrease in drought days, whereas the GCMs project a slight increase in autumn and a decrease in winter. Markers with black edges represent the ensemble mean, and whiskers denote the standard deviation.}
    \label{fig:figure1}
\end{figure}

Emulated and original RCM data agree closely on both the key seasonal patterns and the magnitude and direction of projected changes in drought days, as shown in \Cref{fig:figure1} for daily spatial averages. Our seasonal analysis reveals that drought days are absent in spring and summer across all projections. Overall, the spread of the 80 emulated RCM samples is generally larger than that of the 8 original RCM samples; we examine the spread in more detail in \Cref{sec:sample-size}. Further, the number of drought days in the GCMs is noticeably lower than in the (emulated) RCMs. In \Cref{tab:change-RCM-emulated-GCM} we focus on the changes between the period 2030-39 and 2090-2099, visualized in \Cref{fig:figure1}. Most notably, in autumn and winter, both the original and the emulated RCMs project fewer drought days by the end of the century, whereas the GCMs suggest a slight increase in autumn and a decrease in winter. 

\begin{table}[H]
    \centering
    \small
    \caption{Change of the number of drought days between 2030–2039 and 2090–2099 in seasons with drought days. \textit{Agreement} describes the percentage of models or samples that agree on the sign of \textit{$\Delta$mean}.}
\begin{tabular}{p{2.5cm} p{1.4cm} p{1.6cm} p{2cm} p{1.4cm} p{1.8cm}}
\hline
Dataset & Season & $\Delta$mean [\%] & $\Delta$mean [days]& $\Delta$std [\%] & Agreement $\Delta$mean [\%] \\
\hline
Emulated CMIP5& Autumn & -20.4 &-2.4& -15.2 & 70.0 \\
 (80 samples) & Winter & -25.5 &-10.7& -41.7 & 82.5 \\ \hline
RCM CMIP5 & Autumn & -30.7 &-3.4& -51.0 & 75.0\\
 (8 models) & Winter & -23.9 &-8.5& -16.9 & 100.0 \\\hline
GCM CMIP5& Autumn & +133.3 &+4.0& 0.0 & 100.0\\
 (3 models) & Winter & -30.6 &-6.3& -28.1 & 100.0\\\hline
 Emulated CMIP6 & Autumn & -41.2& -5.0&-10.5& 88.8\\
 (80 samples) & Winter &-11.8 &-3.7&-21.7& 62.5\\\hline
  GCM CMIP6 & Autumn &-45.5&-1.7&-50.5&33.3 \\
 (3 models) & Winter &-18.2&-2.0&+9.1& 100 \\
\hline
\end{tabular}
    \label{tab:change-RCM-emulated-GCM}
\end{table}

\begin{figure}[htbp]
    \centering
    \begin{minipage}[t]{0.48\textwidth}
        \centering
        \includegraphics[width=\linewidth]{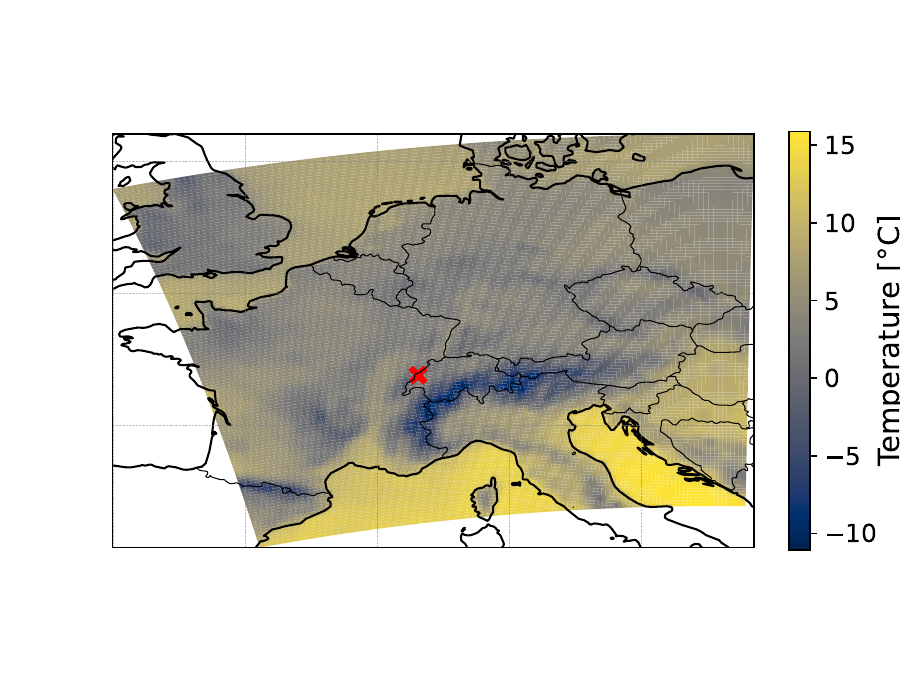}
        \caption{The colored region corresponds to the RCM region considered, and the frame of the plot marks the GCM region. The red marker indicates the RCM location (lon = $6.55^\circ$E, lat = $46.90^\circ$N) analyzed in \Cref{fig:single-location-eraland}.}
        \label{fig:location}
    \end{minipage}\hfill
    \begin{minipage}[t]{0.48\textwidth}
        \centering
        \includegraphics[width=\linewidth]{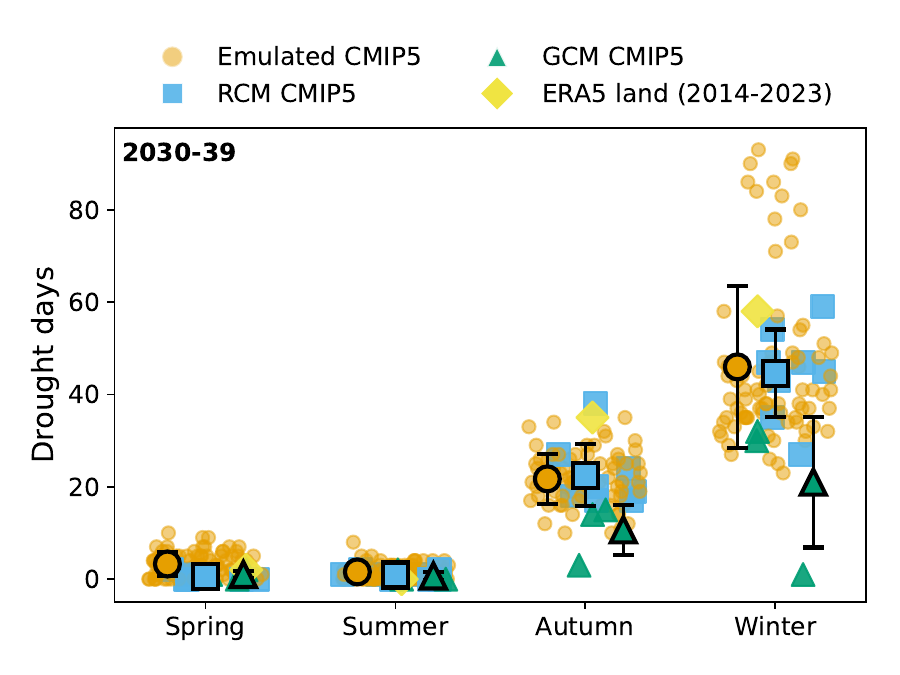}
        \caption{Drought days at a single location. For the RCM and emulated RCM, the location is lon = $6.55^\circ$E, lat = $46.90^\circ$N. For the GCM, the closest grid point to that is lon = $6.25^\circ$E, lat = $46.25^\circ$N and for ERA5-Land lon = $6.50^\circ$E, lat = $46.90^\circ$N. }
        \label{fig:single-location-eraland}
    \end{minipage}
\end{figure}

To complement the regional assessment, we examine a single randomly selected location, which is shown in \Cref{fig:location}. At this location, the change in the number of drought days is very similar between the original and emulated RCMs, providing support that the emulator captures local droughts in a manner similar to that of the original RCM. Comparisons with ERA5-Land for 2014–2023 show that the observed variability falls within the range of RCM simulations, but not within that of GCMs, as illustrated in \Cref{fig:single-location-eraland}. Although ERA5-Land excludes sea areas and, hence, does not cover the full region, the comparison of the entire region, shown in \Cref{fig:appendix-era5land}, provides a qualitative indication that ERA5-Land aligns more closely with the original and emulated RCMs than with the GCM data. ERA5-Land shows more drought days than most regional models, likely because it contains land-only data; regions over the sea, where drought days are fewer (see \Cref{fig:spatial}), are absent. We note that this experiment compares slightly different quantities and should be interpreted cautiously.

\newpage
\subsubsection{Spatial consistency}
\begin{figure}[H]
    \centering
    \includegraphics[width=\linewidth]{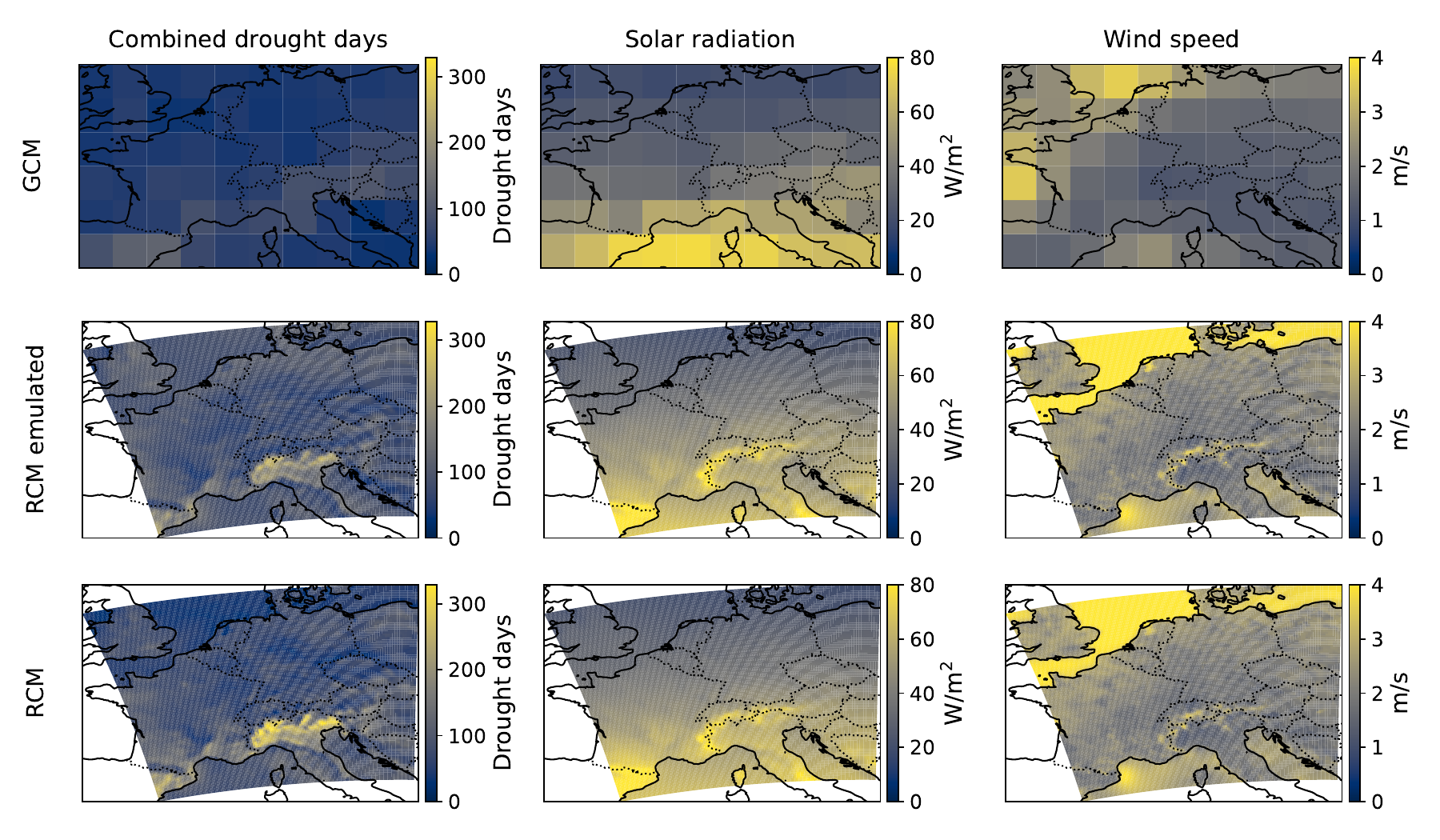}
    \caption{Spatial distribution of co-occurring low wind speed and low solar radiation events in 2090-99 in the CNRM-CM5  global model downscaled with the regional CM5 RegCM4-6 model. The first column shows the number of drought events with location-specific threshold for GCM, RCM, and ML emulator data, illustrating the added spatial detail in RCM and emulator outputs. The second and third columns display the corresponding average solar radiation and average wind speed during drought events, respectively.}
    \label{fig:spatial}
\end{figure}

To assess spatial consistency, we also evaluate the co-occurrence of low wind speeds and low solar radiation at each location (see \Cref{fig:spatial}). For this purpose, we compute location-specific thresholds as described in \Cref{eq:threshold}, \ie the thresholds per location differ. The spatial distribution of drought events is broadly similar between the GCM and RCM (first column); yet, the finer spatial structures introduced by the RCM and the ML emulator are clearly visible. Notably, RCM and emulated RCM show very similar patterns, with only small changes in magnitude. Comparable results are observed for the average solar radiation (second column) and average wind speed (third column) during drought events. Further, the GCM data underestimate location-specific drought occurrences, with the maximum number of drought days being less than half of that found in the RCM and emulated RCM data. We find qualitatively similar results for all other models (not visualized).

\subsection{Application to CMIP6}
\begin{figure}[H]
    \centering
    \includegraphics[width=\linewidth]{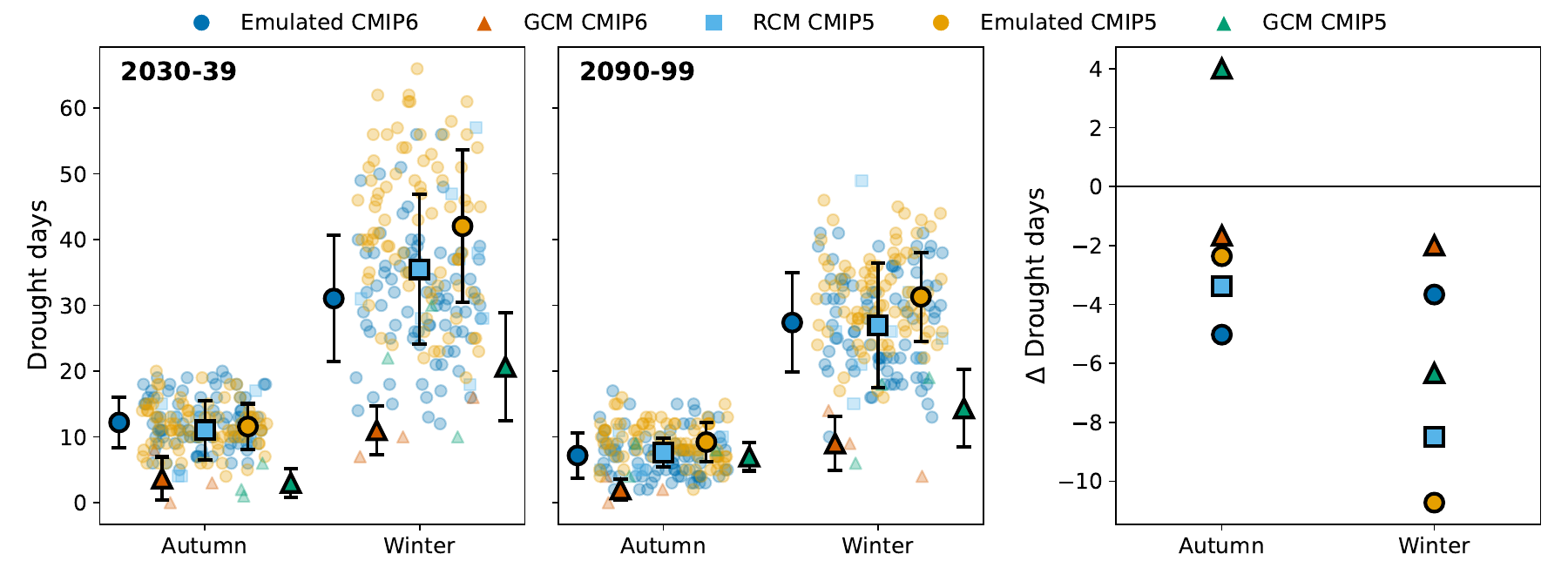}
        \caption{Number of co-occurring low wind and solar days in the original RCMs, emulated RCMs, and GCMs for the periods 2030–2039 (a) and 2090–2099 (b) and the difference between the two time periods (c). The plot is an extension of \Cref{fig:figure1} and additionally includes CMIP6 results. Spring and summer have been omitted as no drought days occur in any of the datasets.}
    \label{fig:cmip6-drought}
\end{figure}

After validating the emulator on CMIP5 and CORDEX data, we apply it to CMIP6 simulations. Examining the spatial plots reveals patterns that are qualitatively consistent with those obtained for CMIP5 (not shown).
The analysis highlights two key findings. First, as with CMIP5, downscaling to regional resolution increases the number of identified drought days compared to the underlying GCMs. Second, despite differences in absolute numbers, the general trend -- a decrease in drought days toward the end of the century -- is consistent across the original RCMs, emulated RCMs, and GCMs in both CMIP5 and CMIP6 (\Cref{fig:cmip6-drought}). Overall, the GCMs show fewer drought days in CMIP6 than in CMIP5, and this pattern is likewise reflected in the RCM data.
Quantitatively, in CMIP5, the expected mean difference in drought days across all seasons over the 10-year period is $-13.1$ for the emulated RCM, $-11.9$ for the original RCM, and $-2.3$ for the GCM. In CMIP6, the corresponding values are $-8.7$ for the emulated RCM and $-3.7$ for the GCM. Overall, results from both CMIP5 and CMIP6 indicate that changes in compound drought days are more pronounced at higher spatial resolutions. 

\subsection{Sample size sensitivity and robustness across ML models}
\label{sec:sample-size}
\begin{figure}[H]
    \centering
    \begin{minipage}[t]{1\textwidth}
        \centering
        \includegraphics[width=0.5\linewidth]{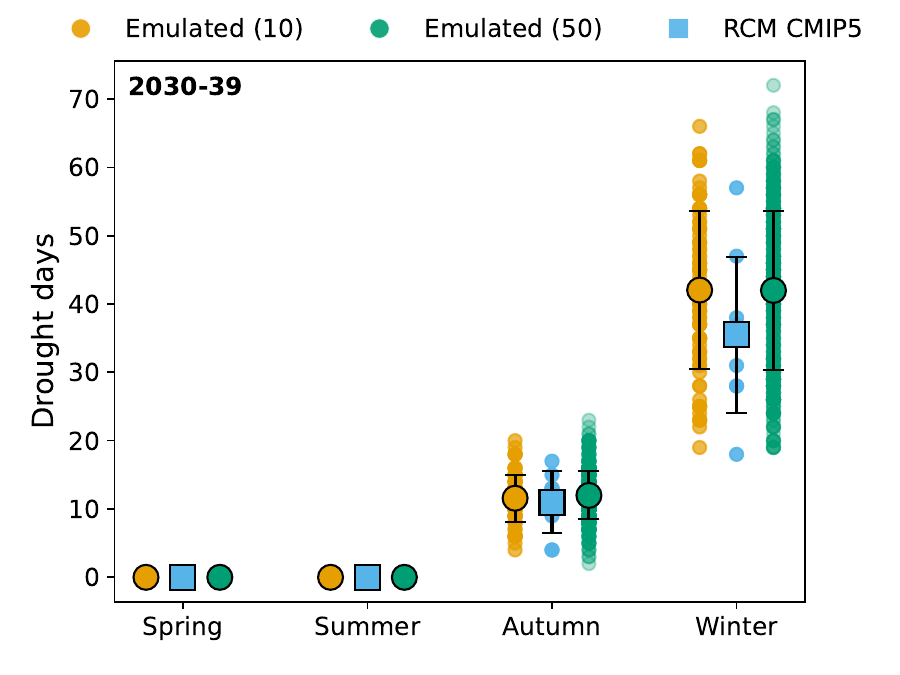}
        \caption{Comparison of 10 vs 50 emulated samples per model. The data stems from emulated CMIP5 and original CMIP5 RCM data and covers the period 2030-2039. Mean and variance are consistent across both sample sizes. Drawing more samples, however, yields higher extremes.}
        \label{fig:moresamples}
    \end{minipage}
\end{figure}

To evaluate the effect of sample size, we compare results based on 10 samples (the samples also used for the other experiments) with those from 50 samples and present them in \Cref{fig:moresamples}. The mean number of drought days per season differs only slightly ($-0.06$ in autumn and $+0.41$ in winter) when using 50 instead of 10 samples.  

The differences in standard deviation are also small ($+0.10$ in autumn and $+0.05$), indicating that both sample sizes capture the mean and standard deviation consistently. As expected, considering more samples increases the likelihood of more extreme values, \ie more drought days: the maximum number of drought days rises from 20 (10 samples) to 23 (50 samples) in autumn and from 66 to 72 in winter. In both 10 and 50 samples, more and higher extremes appear than in the original RCM data, which essentially represent a single realization (one sample).

To assess the robustness across ML models, we trained a second ML model, \textit{EnScale} \citep{schillinger2025enscaletemporallyconsistentmultivariategenerative}, on the same CMIP5–CORDEX pairs. The trends in CMIP5 and CMIP6, as well as the spatial patterns, closely match those obtained with \textit{CorrDiff} (see \Cref{fig:enscale-cmip-cmip6} and \Cref{fig:spatial-enscale}). This demonstrates that the main conclusions regarding the frequency and spatial distribution of co-occurring wind and solar droughts are not dependent on our specific choice of ML architecture.

\section{Discussion}
\label{sec:discussion}
Our results indicate a decrease in the number of co-occurring wind speed and solar radiation drought days toward the end of the century. 
However, both the original RCMs and the emulated RCMs consistently exhibit more compound drought days than the driving GCMs in both CMIP5 and CMIP6, indicating that regional models capture extremes that are smoothed out in coarse global projections. Comparison with ERA5-Land indicates that GCMs may underestimate the frequency of drought days, whereas RCMs and the emulator capture high-resolution variability that is more consistent with reanalysis data. This supports previous findings that higher resolution can improve the representation of extremes \citep{morelli2025climate, byun2023investigation}. Therefore, our results highlight the added value of RCMs for our specific research question, as they more accurately represent compound drought events than their driving GCMs. 
The decrease in compound drought events likely stems from a seasonal shift in wind drought events. This shift is visualized in \Cref{fig:wind-drought}, which shows a rise in summer wind drought events, broadly consistent with \citet{reyers2016future}, who find reductions in European summer wind speeds. The emulator successfully reproduces the signal of the CORDEX data, demonstrating its ability to capture regional patterns. When applied to CMIP6, the overall change in drought days is slightly smaller than in CORDEX, reflecting differences in the underlying global models that are consistent with other research \citep{carvalho2021wind, ha2023comparative}.

The consistency of our results across both GCM and RCM datasets, including the lower number of overall drought days in (emulated) CMIP6 compared to (emulated and RCM) CMIP5 in \Cref{fig:cmip6-drought}, suggests that the emulator has successfully learned the underlying statistical relationships between large-scale climate fields and regional extremes, enabling it to generalize beyond the CMIP5 training data.
Our findings show that the ML emulator can downscale GCM projections to RCM resolution at a small fraction of the computational cost of an RCM while retaining characteristics relevant to energy systems. A key benefit is that such emulators can be applied to new climate model generations where RCM simulations are still lacking, including CMIP6, while avoiding both the substantial computational expense of regional modeling and the coordination demands of initiatives such as CORDEX.
A sensitivity analysis with larger sample sizes further supports the robustness of our findings. Comparing 10 samples (used in the main experiments) with 50 samples revealed only marginal differences in the mean and variability of drought days, while a larger number of samples primarily increased the chance of capturing extreme values. This indicates that our conclusions are not sensitive to sample size. 

The results presented can also be partially contrasted with the study by \citet{strnad2025assessingriskfuturedunkelflaute}, which employed a machine learning approach to relate CMIP6 simulations directly to ERA5 reanalysis data to assess future energy drought events in Germany. However, the approach in their study is fundamentally different: our emulator is trained on high-resolution CORDEX RCM data and CMIP5 GCMs, learning the statistical mapping between regional and global models, whereas \citet{strnad2025assessingriskfuturedunkelflaute} train their model on historical observations and apply it to CMIP6. They find no significant changes in the occurrence of energy drought events, whereas our results suggest a slight decrease toward the end of the century, highlighting how training data, model setup, and the definition of a drought event can influence projections of future extremes.

Our results also emphasize that the overall skill of ML emulators does not necessarily translate into the best performance in impact-relevant scores. Previous studies identified the best general performance with larger machine learning architectures than those used here \citep{mardani2025residual,schillinger2025enscaletemporallyconsistentmultivariategenerative}. However, in our analysis, a smaller architecture showed benefits for predicting the specific case of compound energy droughts (see \Cref{sec:additional_plots_spatial_avg}), and we find that the more complex model performs worse in predicting seasonal wind droughts (see \Cref{fig:wind-attention}).

Drought days per year and per season are rare events, so interpreting the resulting climate signals requires particular caution. Diverging trends between GCMs and RCMs -- for example, in autumn -- show that even when emulated datasets reproduce RCM trends and variability well, comparisons across model resolutions must be approached carefully. Uneven sample sizes among GCMs, RCMs, and emulated data also complicate the comparison of mean changes reported in \Cref{tab:change-RCM-emulated-GCM}. Following the recommendation of \citet{von2013testing}, we refrain from applying formal statistical tests. While such tests are often employed to assess significant changes, uneven ensemble sizes make their interpretation problematic in our analysis \citep[\eg][]{serdar2021sample}. For instance, in CMIP5, a paired t-test at the 0.05 level with Bonferroni correction would deem the changes in the original RCM and GCM insignificant, whereas smaller changes in the emulated datasets would appear significant. This contrast illustrates a limitation of unequal sample sizes but also highlights a strength of probabilistic machine learning approaches: the large number of emulated realizations enables more robust statistical assessments if desired. While large ensembles of single-model climate simulations \citep{kay2015community} could also be used for such analyses, they are computationally expensive and rather rarely available.

We demonstrate the utility of ML emulators using a drought-day definition from related literature. We select this definition because it avoids the need for bias correction and does not require transforming wind speed or solar radiation into capacity factors.
We identify droughts based on daily data, which may not fully capture short-term or local variations in conditions. Using daily data also limits the transformation to capacity factors, as a finer temporal resolution is required to accurately reflect power generation \citep{effenberger2023mind}. Additionally, our approach does not account for electricity demand, which can significantly influence the characterization and severity of drought days \citep{kittel2024measuring}. Finally, we analyze wind and solar time series separately rather than in combination, since combining them would require assuming the relative importance of each energy source. This drought definition was chosen deliberately to avoid imposing such assumptions.

\subsection{Future work}
As a result, while the current methodology provides a useful overview, it may overlook critical aspects relevant to operational planning and impact assessments. Therefore, future work should incorporate more elaborate definitions of compound energy droughts \citep{kittel2024measuring}, including capacity factors, locations of renewable power installations \citep{effenberger2025turbine}, and potentially also electricity demand \citep{verwiebe2021modeling, grochowicz2024using}. Furthermore, in real-world applications, the duration of Dunkelflaute events is crucial. \textit{CorrDiff}, however, does not ensure physical consistency \citep{mardani2025residual}. Alternative approaches, such as \textit{EnScale-t} \citep{schillinger2025enscaletemporallyconsistentmultivariategenerative}, show promise in this regard and could be exploited in future work. Re-training a machine learning model with additional RCM-GCM pairs and capacity factors directly could further align the downscaling approach with energy system applications. Additionally, energy system optimization methods \citep{de2025climate} could potentially benefit from climate model emulators and large sample sizes. \\

Since the number of samples can be increased arbitrarily, it becomes possible to investigate whether this scalability enables estimating return periods \citep{bloin2025estimating} and extending the analysis toward rare extremes using specialized methods \citep{pandey2024heavy}. This is relevant not only for renewable-energy applications but also for broader uncertainty and variability assessments in other domains. The generative framework of ML models, such as \textit{CorrDiff} and \textit{EnScale}, provides a basis for quantifying uncertainties.

\section{Conclusion} 
In this study, we demonstrate that machine learning emulators can effectively downscale coarse global climate model projections to the resolution of regional climate models, capturing key characteristics of drought events relevant to energy systems. The emulator reproduces RCM signals and variability, and our analysis highlights the added value of high-resolution modeling by revealing more frequent droughts in high-resolution data than indicated by the driving GCMs. When applied to CMIP6 data, the emulator indicates slightly fewer changes in the number of drought days compared to CORDEX, reflecting differences in the underlying global models.
Our results emphasize the importance of validating emulators on impact-relevant metrics, and the probabilistic nature of the emulator enables the generation of large ensembles, facilitating robust statistical analyses. Overall, our study highlights the potential of machine learning emulators as a computationally efficient tool to bridge global and regional climate projections, supporting more detailed insights for impact studies.

\subsection*{Acknowledgements}
Nina Effenberger, Maybritt Schillinger, Maxim Samarin, and Reto Knutti are part of SPEED2ZERO, a Joint Initiative co-financed by the ETH Board. 

\newpage
\bibliographystyle{plainnat}
\bibliography{bib}
\newpage
\newpage
\begin{appendices}

\counterwithin{figure}{section}
\counterwithin{table}{section}
\section{Additional plots for spatial averages}
\label{sec:additional_plots_spatial_avg}

In the main part of the paper, we use a smaller U-Net architecture with about 10 million parameters (see Section \ref{paragraph:ml_model}) compared to the original setting in \citet{mardani2025residual} with about 80 million parameters. In the following, we refer to the U-Net architecture in this work as ``no attention'', while ``with attention'' corresponds to the original setting of \citet{mardani2025residual} with an adapted self-attention layer at the bottleneck representation with feature map resolutions of $8\times 8$ (as our input is of image size $128 \times 128$). Our choice is motivated by the results in Figure \ref{fig:attention-emulator-cmip5}: ``no attention'' samples generally agree better with the RCM results compared to the ``with attention'' model.

\begin{figure}[H]
    \centering
    \includegraphics[width=\linewidth]{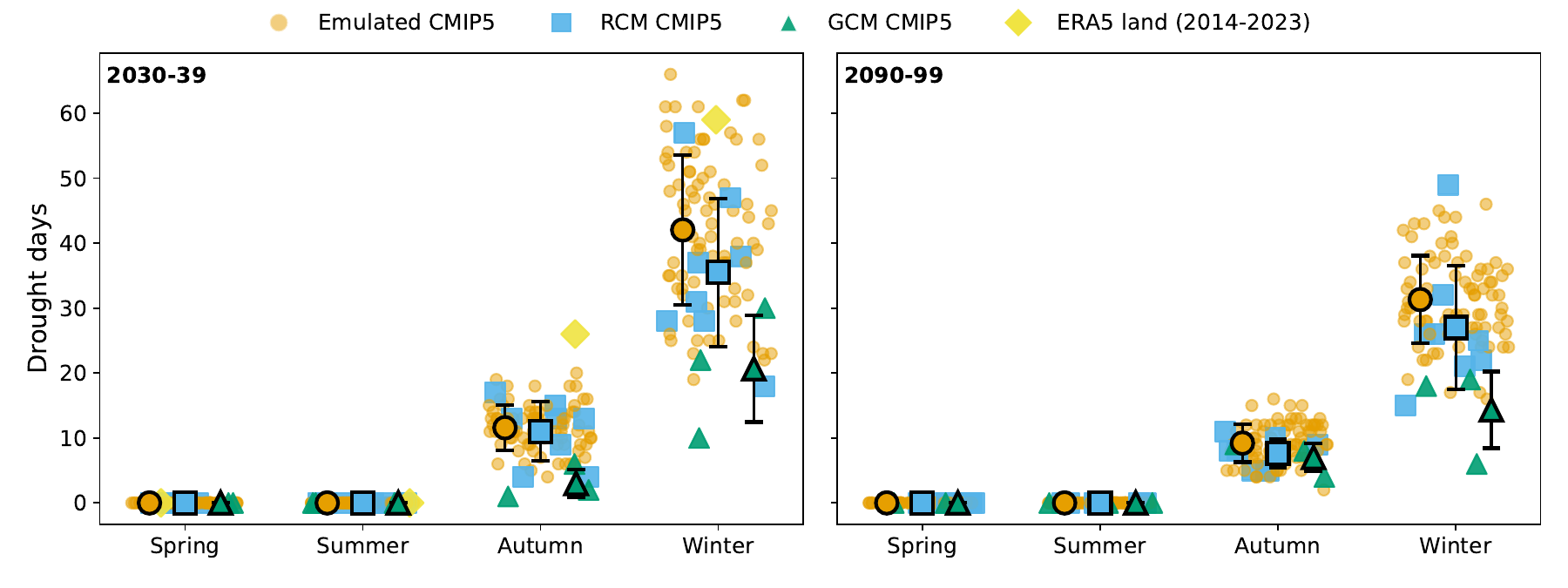}
    \caption{Comparison of drought days between CMIP5 and ERA5-Land across the entire study region. ERA5-Land spans the same area as the GCM data at approximately RCM resolution. Since ERA5-Land is undefined over the ocean, these results should be interpreted with caution. }
    \label{fig:appendix-era5land}
\end{figure}

\begin{figure}[H]
    \centering
    \includegraphics[width=\linewidth]{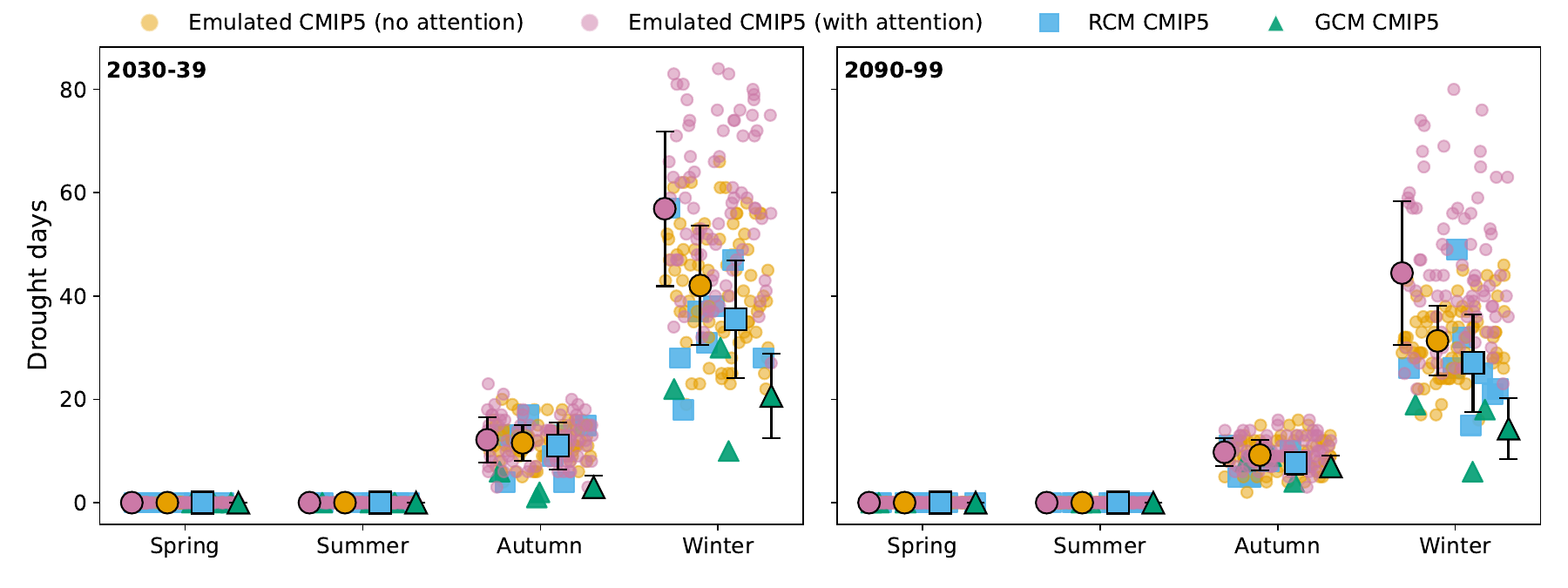}
    \caption{Comparison of CorrDiff models on CMIP5 data using different U-Net architectures. The (larger) model with self-attention tends to overestimate winter droughts compared to RCMs, which is why we use the (smaller) version without self-attention for this study.}
    \label{fig:attention-emulator-cmip5}
\end{figure}

\begin{figure}[H]
    \centering
    \includegraphics[width=\linewidth]{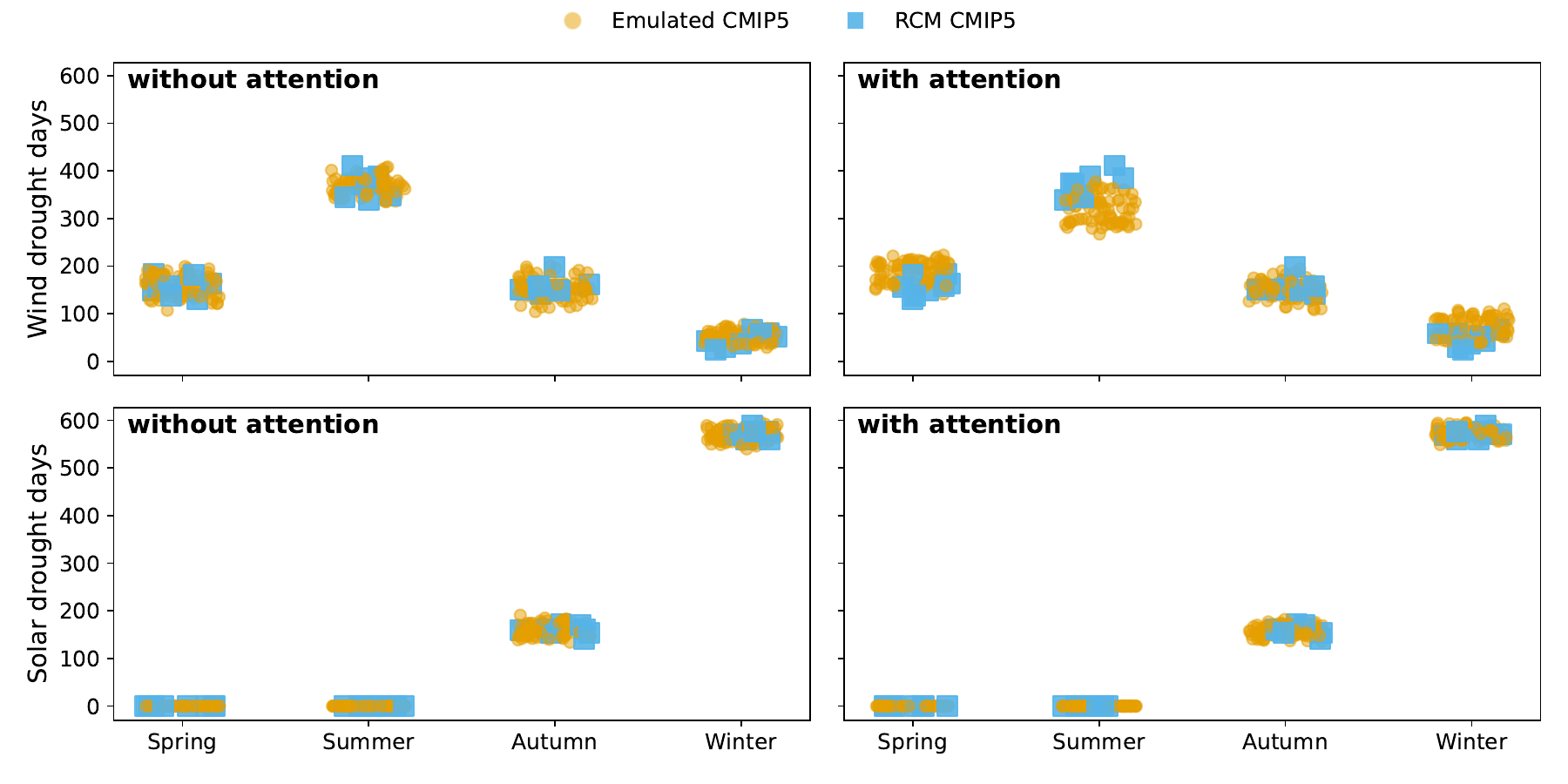}
    \caption{Seasonal counts of solar and wind drought days in emulators with and without attention. The attention-based emulator produces more summer wind drought days than both the original RCM data and the non-attention emulator.}
    \label{fig:wind-attention}
\end{figure}

\begin{figure}[H]
    \centering
    \includegraphics[width=\linewidth]{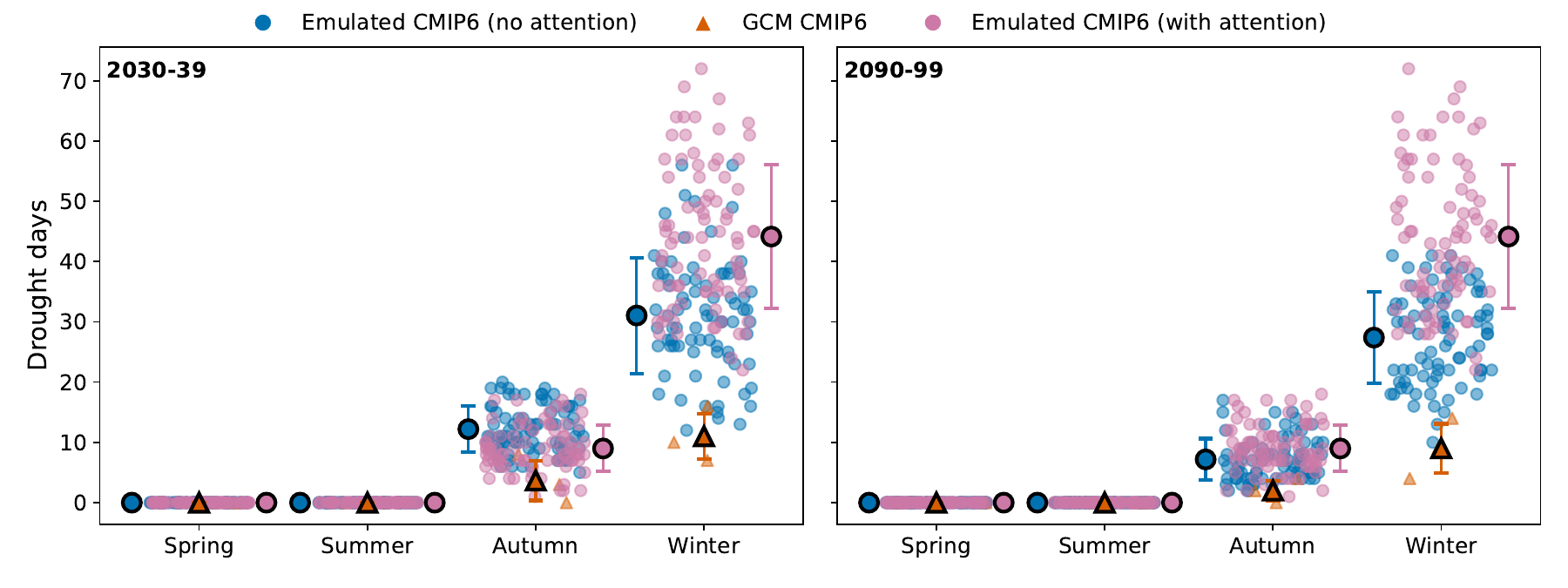}
        \caption{Comparison of CorrDiff models on CMIP6 data using different U-Net architectures. As in the CMIP5 case, the (larger) model with self-attention projects a higher number of drought days than the (smaller) model without self-attention.}
    \label{fig:attention-emulator-cmip6}
\end{figure}

\begin{figure}[htbp]
    \centering
    \begin{minipage}[t]{0.48\textwidth}
        \centering
        \includegraphics[width=\linewidth]{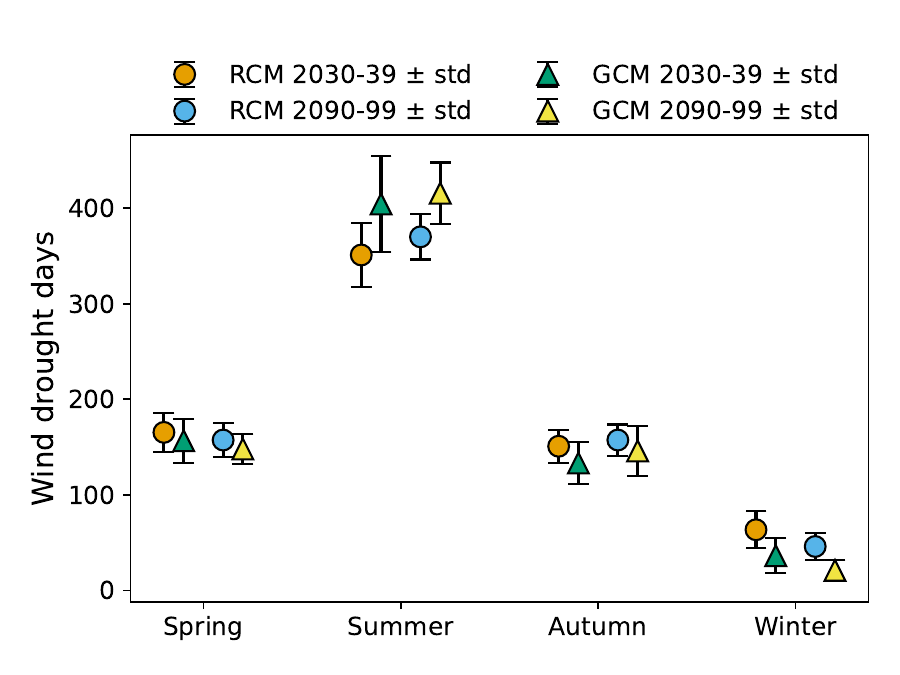}
        \caption{Wind drought days in the original GCM and RCM data across the whole study region. In both periods, the GCM exhibits more summer drought days and fewer winter drought days than the RCM. Both, RCMs and GCMs, project a slight increase in summer drought days and a slight decrease in winter drought days. }
        \label{fig:wind-drought}
    \end{minipage}\hfill
    \begin{minipage}[t]{0.48\textwidth}
        \centering
        \includegraphics[width=\linewidth]{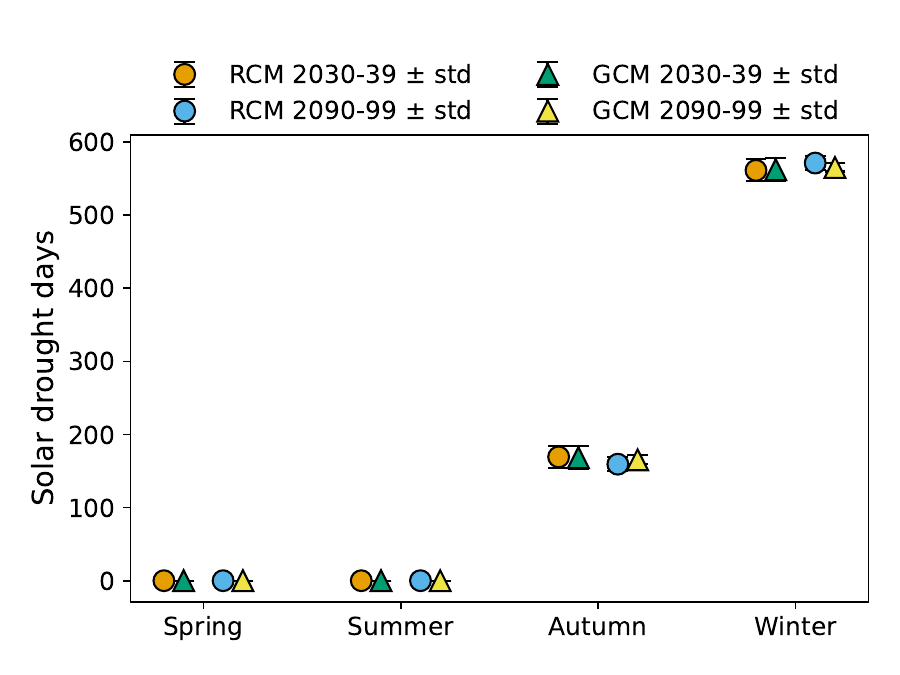}
        \caption{Solar drought days in the original GCM and RCM data. The seasonal number of solar drought days remains relatively stable across models and time periods.}
        \label{fig:solar-drought}
    \end{minipage}
\end{figure}
\newpage
\section{Results for EnScale}
\label{sec:enscale}
In the main part of the paper, we focus on the machine learning model \textit{CorrDiff} by \citet{mardani2025residual}. Here, we present the results using another model, \textit{EnScale} of \citet{schillinger2025enscaletemporallyconsistentmultivariategenerative}.
\begin{figure}[H]
    \centering
    \includegraphics[width=\linewidth]{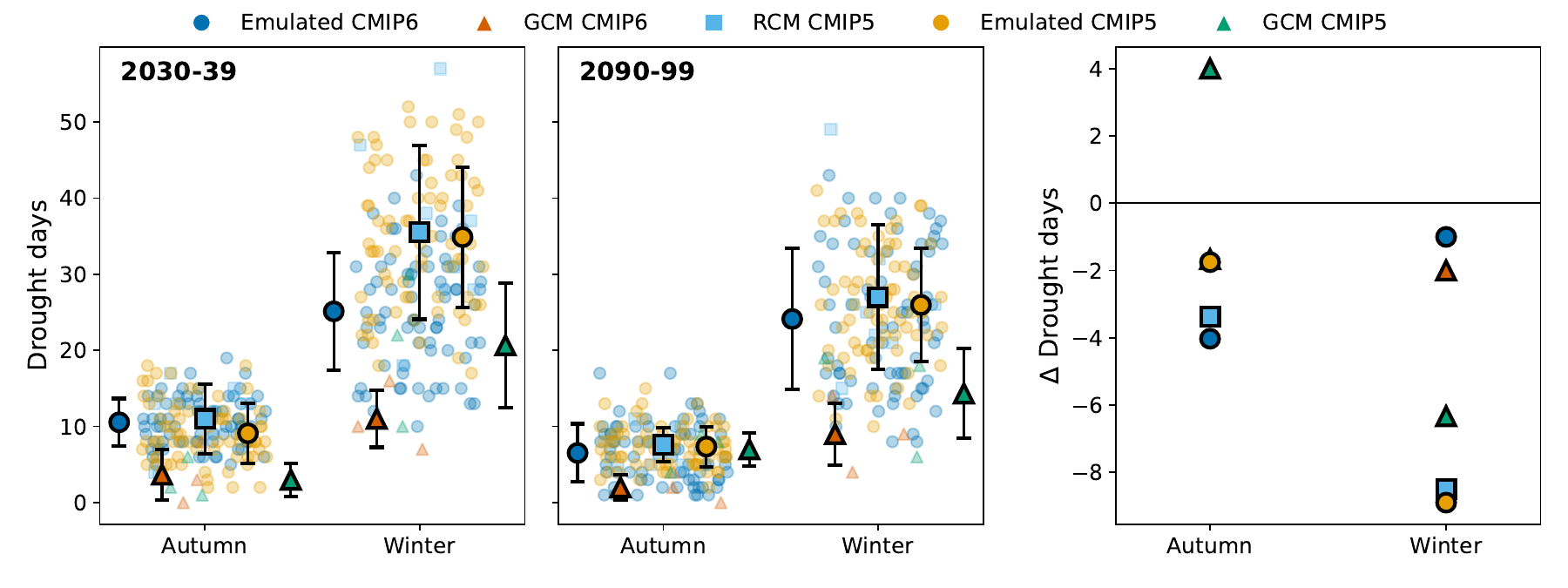}
        \caption{Number of co-occurring low wind and solar days in the original RCMs, emulated RCMs, and GCMs for the periods 2030–2039 (a) and 2090–2099 (b) and the difference between the two time periods (c). Same plot as \Cref{fig:cmip6-drought} but for the ML model \textit{EnScale}. }
    \label{fig:enscale-cmip-cmip6}
\end{figure}
\begin{figure}[H]
    \centering
    \includegraphics[width=\linewidth]{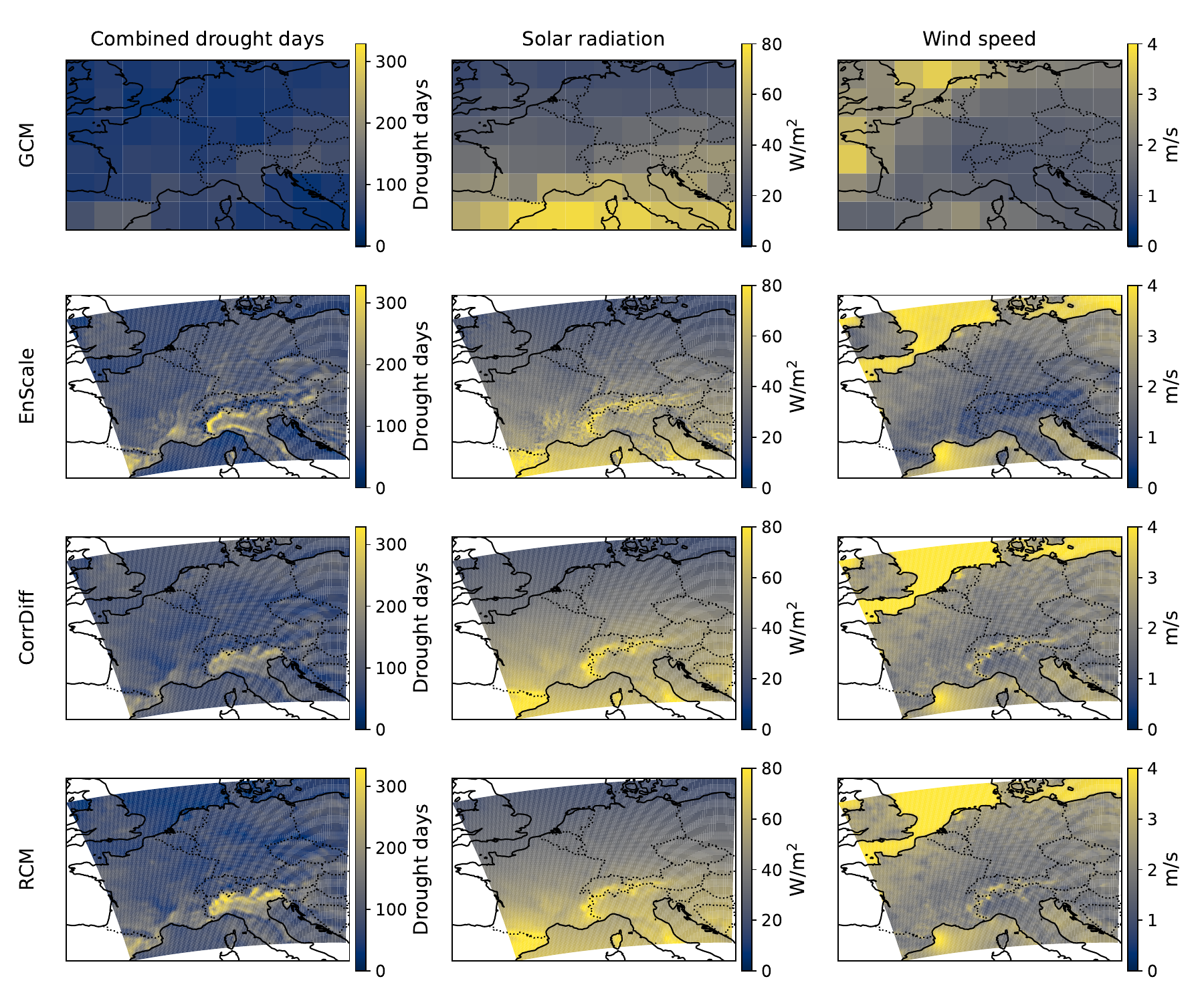}
        \caption{Spatial distribution of co-occurring low wind speed and low solar radiation events in the CNRM global model downscaled with the regional CM5 RegCM4-6 model. The first column shows the locations of drought events with location-specific threshold for GCM, RCM, and ML emulator data, illustrating the added spatial detail in RCM and emulator outputs. The second and third columns display the corresponding average solar radiation and average wind speed during drought events, respectively. Same plot as \Cref{fig:spatial} but with \textit{EnScale} added.}
    \label{fig:spatial-enscale}
\end{figure}

\end{appendices}
\end{document}